\documentclass[sigconf]{acmart}
\AtBeginDocument{%
  }

\setcopyright{acmlicensed}
\copyrightyear{2018}
\acmYear{2018}
\acmDOI{XXXXXXX.XXXXXXX}

\acmConference[Conference acronym 'XX]{Make sure to enter the correct
  conference title from your rights confirmation email}{June 03--05,
  2018}{Woodstock, NY}

\author{Jingjia Xiao}
\affiliation{
  \institution{Department of Sociology, University of California San Diego}
  \city{La Jolla}
  \state{California}
  \country{USA}
}
\email{jix082@ucsd.edu}

\author{Qing Xiao}
\affiliation{
  \institution{Human-Computer Interaction Institute, Carnegie Mellon University}
  \city{Pittsburgh}
  \state{Pennsylvania}
  \country{USA}
}
\email{qingx@cs.cmu.edu}

\author{Hong Shen}
\affiliation{
  \institution{Human-Computer Interaction Institute, Carnegie Mellon University}
  \city{Pittsburgh}
  \state{Pennsylvania}
  \country{USA}
}
\email{hongs@cs.cmu.edu}

\acmISBN{978-1-4503-XXXX-X/2018/06}

\usepackage{enumitem}
\usepackage{longtable}
\usepackage{booktabs}
\usepackage{tabularx}
\usepackage[table]{xcolor} 

\AtBeginDocument{
  }

\begin{document}

\title{“It Felt Real” Victim Perspectives on Platform Design and Longer-Running Scams}

\begin{abstract}
Longer-running scams, such as romance fraud and “pig-butchering” scams, exploit not only victims’ emotions but also the design of digital platforms. Scammers commonly leverage features such as professional-looking profile verification, algorithmic recommendations that reinforce contact, integrated payment systems, and private chat affordances to gradually establish trust and dependency with victims. Prior work in HCI and criminology has examined online scams through the lenses of detection mechanisms, threat modeling, and user-level vulnerabilities. However, less attention has been paid to how platform design itself enables longer-running scams. To address this gap, we conducted in-depth interviews with 25 longer-running scam victims in China. Our findings show how scammers strategically use platform affordances to stage credibility, orchestrate intimacy, and sustain coercion with victims. By analyzing scams as socio-technical projects, we highlight how platform design can be exploited in longer-running scams, and point to redesigning future platforms to better protect users.
\end{abstract}



\maketitle

\section{Introduction}
In recent years, as digital platforms have become deeply embedded in social, consumer, and financial life, online scams have taken on new forms of evolution \cite{kraemer2013cybercrime,tambe2024towards,brogi2024new,scharfman2025cryptocurrency,whitty2016online,carter2021distort}. Among them, longer-running scams, such as “pig-butchering” and romance fraud, have proven especially damaging \cite{agarwal2025overview,whitty2016online}. Unlike short-term phishing attacks that aim for immediate gains, these scams unfold over weeks or even months, gradually cultivating victims’ trust and dependence \cite{scamwatchhq2025pig}. They not only manipulate victims’ emotional vulnerabilities but also systematically exploit platform design features: polished identity verification mechanisms, persistent algorithmic recommendations, embedded payment systems, and relatively concealed private chat functions \cite{perdana2024crypto}. The result is harm that extends beyond financial loss, leaving victims with profound psychological trauma and an erosion of social trust \cite{scharfman2025cryptocurrency,button2014not,whitty2016online}. This raises a pressing question:\textit{ what role do platforms, or more specifically, platform design, play in enabling these longer-running scams?}

Yet existing scholarship tends to focus on scams after the fact, analyzing manipulative techniques, information flows, or victims’ vulnerabilities in emotion, trust, and risk assessment \cite{norris2019psychology}. A critical but overlooked question concerns how victims are first drawn into scams. In other words, the “access” point, i.e., how initial contact between victims and scammers is established, is rarely accidental. Instead, according to many news and police statement, it is often embedded within platform mechanisms such as recommendation algorithms, group features, friend expansion tools, or advertising distribution systems \cite{lanzhou2025safewatchscam,arong2025sextortion,jiyuan2025targetedfraud}. Scammers are not isolated actors; they strategically leverage platform affordances to funnel potential targets into private, less visible spaces of interaction. Notably, these spaces win victims’ trust precisely because they mimic or repurpose the design logic of legitimate platforms. Similar interface appearances, seemingly professional identity badges, and interaction flows consistent with mainstream social platforms together create a sense of familiar safety \cite{lanzhou2025safewatchscam,arong2025sextortion,jiyuan2025targetedfraud}. Victims often do not perceive themselves as stepping outside a legitimate platform ecosystem, but rather believe they remain within it. This simulated design trust further reduces vigilance and allows scams to continue undetected over extended periods.

To address this gap, we conceptualize scams as socio-technical processes. They involve not only scammers’ manipulation of emotion and relationships but also platform design features that facilitate both access and trust-building. 

Therefore, in this paper, we ask two guiding questions: 
\begin{itemize}
    \item \textbf{RQ1: }How are victims gradually funneled into scams through platform mechanisms?
    \item \textbf{RQ2: }How do hard-to-monitor interaction spaces leverage the design logics of legitimate platforms to gain and sustain trust?
\end{itemize}
Answering these questions moves us beyond narrow emphases on individual vulnerability or detection technology, toward understanding scams as socio-technical projects embedded in platform infrastructures.

To investigate this dynamic, we conducted semi-structured interviews with 25 victims in China who had experienced longer-running scams such as pig-butchering and romance fraud. China provides a particularly important site for studying these dynamics. The country has witnessed a surge of longer-running scams such as pig-butchering and romance fraud, many of which explicitly target middle-class aspirations for financial mobility and exploit the popularity of mainstream social platforms like Xiaohongshu and WeChat \cite{lanzhou2025safewatchscam,arong2025sextortion,jiyuan2025targetedfraud}. At the same time, China’s rapid platformization of financial and social life—where payment, investment, and communication are tightly integrated—creates unique conditions for scammers to appropriate platform features in ways that blur the boundary between legitimate and fraudulent practices \cite{lanzhou2025safewatchscam,arong2025sextortion,jiyuan2025targetedfraud}. Victims’ stories in this context thus offer crucial insights into how platform affordances can be strategically misused.  

Our findings reveal a layered process through which scams take root and persist. First, entry is rarely accidental: victims are funneled into scams through mainstream platforms, where recommendation systems and interaction features amplify exposure and normalize fraudulent content. Second, scammers fabricate legitimacy by strategically appropriating platform design elements—such as trading interfaces, embedded payment flows, and customer service chats—to simulate the look and feel of formal institutions. Finally, scams are sustained through the social dynamics of platforms: group interactions, emotional support, and disciplinary mechanisms bind victims into networks that are difficult to leave. Together, these dynamics demonstrate how scams exploit the very affordances that underpin everyday platform use, transforming them into infrastructures for deception and control.

This study thus makes two contributions to HCI, online safety and criminology community:
\begin{itemize}
    \item This study provides rare, in-depth qualitative material from interviews with 25 victims, revealing how scams unfold longitudinally. It advances the concept of scams as socio-technical projects, highlighting the systemic role of platforms in enabling both “access” and “trust-building” and thereby expanding HCI’s dominant explanatory frameworks of online fraud.
    \item We offer implications for platform design and governance, emphasizing the need to critically reassess identity verification mechanisms, recommendation algorithms, payment integrations, and private chat affordances to prevent their strategic misuse. By foregrounding the deep entanglement of scams and platform design, we call for future safety design to move beyond user education and detection tools, embedding scam prevention into the very structures of platforms themselves.
\end{itemize}

\section{Related Work}
Online scams have been studied across multiple disciplines, including criminology, online safety, and Human–Computer Interaction. This body of work has traced how scams evolve alongside technological change (\autoref{lr1}), how vulnerabilities are unequally distributed across different social groups (\autoref{lr2}), and how platform infrastructures and design features can be strategically exploited by scammers (\autoref{lr3}). More recent HCI research has emphasized socio-technical perspectives, moving beyond individual-level explanations toward examining scams as embedded in larger cultural, economic, and technological systems (\autoref{lr4}). Building on these strands of scholarship, our study contributes by analyzing how platform design is appropriated across both the access and trust-building stages of longer-running scams.

\subsection{Typologies and Evolution of Scams} \label{lr1}

Early research on scams largely focused on short-term, one-off frauds such as the “Nigerian prince” email scheme or traditional phishing \cite{obiaya2022nigerian,okosun2023evolution,sheng2007anti}. These scams typically relied on email or instant messaging, luring users into clicking malicious links or attachments through false promises of financial gain or fabricated security alerts \cite{obiaya2022nigerian,okosun2023evolution}. Prior work has shown that phishing succeeds in part because users, under conditions of high information load, tend to rely on heuristic cues (such as brand logos or icons) rather than systematic reasoning \cite{sheng2007anti}. Later studies found that the rise of mobile devices has exacerbated this problem: the interface constraints and usage habits of smartphones encourage rapid-clicking behaviors, thereby increasing the likelihood that users fall into traps unconsciously \cite{thakur2018innovations,alsharnouby2015phishing}.

With the expansion of social media and mobile payments, scams have evolved into longer-running, relationship-oriented schemes. Prominent examples include pig-butchering scams and romance fraud, where trust is built over weeks or even months of everyday interactions before victims are gradually funneled into investment or money-transfer schemes \cite{agarwal2025overview,whitty2016online}. Research highlights that scammers in these cases act not only as “attackers” but also as “relationship builders,” cultivating emotional bonds through sustained interaction and transforming scams into forms of deep social and psychological manipulation \cite{dixon2022holding}.

The typologies of scams have also diversified. For instance, scholars have documented how fraudulent job advertisements are systematically used for human trafficking recruitment, with scammers imitating the interface and workflow of legitimate hiring platforms to lure job seekers into dangerous situations \cite{moyo2025investigating}. Similar mechanisms appear in cryptocurrency and investment scams, where scammers frequently create fraudulent websites modeled on mainstream financial platforms, offering promises of high returns to attract investors \cite{leukfeldt2019cybercrime,tambe2024towards}. 

These cases illustrate that scams are no longer merely the result of “information asymmetry,” but are continuously updated and iterated by exploiting emerging technologies and platform logics. However, most studies still concentrate on describing scam techniques, with relatively little attention to how platform design itself enables these schemes.

\subsection{Vulnerable Populations and Social Inequalities} \label{lr2}

Another major line of scam research examines differences in susceptibility across social groups. Older adults are often considered highly vulnerable. Deng et al. \cite{deng2025auntie} found that in China, older people frequently rely on younger family members to help identify and guard against online scams, suggesting that protective practices are not merely individual behaviors but are deeply embedded in family relationships and cultural norms. Western studies similarly note that older adults are more likely to be targeted due to cognitive decline, loneliness, and information isolation \cite{o2024defendants,shang2022psychology,carter2025digital}.

The risks faced by adolescents have also gained increasing attention. Drawing on surveys and Instagram data, Alsoubai et al. \cite{alsoubai2022friends} showed that high-risk youth are more likely to overlook potential dangers in digital interactions, such as forming incomplete or unsafe connections with strangers when seeking recognition or engaging in risk-taking behaviors. This suggests that adolescents’ susceptibility to scams stems not only from limited experience but also from psychological characteristics tied to their developmental stage.

Economic and social status likewise shape susceptibility to scams. Vitak et al. \cite{vitak2018knew}, through interviews with low-income families in the United States, found that limited digital literacy, unstable internet access, and reliance on public devices made disadvantaged groups more vulnerable and undermined their ability to respond to scams. These constraints both heightened risk and deepened distrust of digital systems. Gonzales et al. \cite{gonzales2021dis} further demonstrated that race, education, and class significantly affect users’ perceptions of online risks, with racial minorities and less-educated groups more frequently reporting scam victimization.

These studies reveal how scam risks are deeply embedded within broader structures of social inequality. However, much of this scholarship still centers on the limitations or circumstances of individual users, with less attention to how platform design itself may exacerbate or mitigate such inequalities. In other words, why vulnerable groups are “more easily reached,” and how scammers strategically exploit platform features to target them, remain insufficiently understood.

\subsection{Platform Design and Scam Exploitation} \label{lr3}

As platforms have become increasingly central to social and financial life, scholars have begun to examine their role in enabling scams. Liu et al.\cite{liu2022user} highlight that emerging services often face a “cold start problem”: scammers tend to exploit platform features—such as recommendation algorithms or identity verification—at an early stage, when detection mechanisms are still underdeveloped, in order to boost their visibility or credibility.

More broadly, research shows that a variety of platform affordances are strategically repurposed in scams \cite{downs2006decision}. Algorithmic recommendations and friend-expansion functions allow scammers to repeatedly surface in victims’ feeds \cite{ohu2025examination}, creating a sense of familiarity. Identity verification mechanisms and interface symbols foster an atmosphere of legitimacy. Embedded payment systems lower the psychological barriers to financial transfer. Private chats and closed groups provide hidden spaces where scammers can sustain relationships and maintain coercive control over time \cite{carter2021distort}. Design features originally intended to optimize user experience thus become tools of manipulation when appropriated by scammers \cite{downs2006decision}.

Other studies reveal that scammers not only exploit platform functions but also imitate or replicate the design logics of legitimate platforms to manufacture trust. Moyo et al. \cite{moyo2025investigating}, for instance, found that human traffickers posted fraudulent job advertisements that closely mirrored the interface and workflows of legitimate recruitment platforms, leveraging users’ trust in platform norms to deceive them. Similarly, cryptocurrency investment scams often build websites that mimic the look and feel of mainstream exchanges, reproducing familiar visual and interaction patterns to create a false sense of security \cite{leukfeldt2019cybercrime}. Such “simulated design” demonstrates another mechanism of deception: the reproduction of familiar interaction logics to dampen users’ risk awareness.

These studies illustrate the dual role of platform design in scams: shaping the pathways through which victims first encounter scammers, and sustaining perceptions of legitimacy over time. Yet most existing work remains case-based, lacking a systematic analysis of how scammers strategically exploit platform affordances across both the “access” and “trust-building” stages of manipulation.

\subsection{HCI and Socio-Technical Perspectives on Scamming} \label{lr4}

In the field of Human–Computer Interaction (HCI), scams and security issues have long been an important area of inquiry. Early studies primarily emphasized user education and interface cues, focusing on improving digital literacy or designing more effective warnings to help users recognize scams \cite{sheng2007anti,aneke2021help}. However, as scams have evolved, HCI scholars increasingly recognize that scams are not simply the outcome of individual risk judgments, but are deeply embedded within complex socio-technical systems \cite{deng2025auntie,vitak2018knew}.

Recent work has expanded the scope of scam research by highlighting its multi-actor and cultural dimensions. Razaq et al. \cite{razaq2021we} showed that Pakistan’s mobile scam ecosystem involves not only scammers but also intermediaries, family members, and regulators, illustrating that anti-scam practices are shaped through multi-party interactions. Deng et al. \cite{deng2025auntie} emphasized how intergenerational relationships and cultural norms such as filial piety influence anti-scam practices in China. Vitak et al. \cite{vitak2018knew} and Gonzales et al. \cite{gonzales2021dis} further underscored the role of structural inequalities in shaping scam susceptibility, showing that low-income groups, racial minorities, and less-educated populations face disproportionately high risks in digital environments.

While HCI has broadened scam research by advancing socio-technical perspectives, most existing studies still center on user-side vulnerabilities or interventions such as detection and education. Much less attention has been paid to how scammers systematically exploit platform design to enable both access and trust-building. In other words, the role of platforms in the very occurrence of scams remains underexamined. To more fully understand online fraud—especially longer-running scams—we need to conceptualize them as socio-technical projects embedded in platform logics. This perspective provides a crucial new entry point for studying how scams unfold and persist over time.

\section{Method}
This study aims to understand how longer-running scams leverage platform design to facilitate both “access” and “trust-building,” as well as how victims become gradually drawn into scams over extended interactions. Because these research questions require capturing victims’ subjective experiences, emotional fluctuations, and interpretations of platform mechanisms, we adopted a qualitative research approach, collecting empirical data through semi-structured in-depth interviews. This method allowed participants to reconstruct their scam experiences in their own words, thereby revealing the temporal unfolding of scams, key turning points, and the role of platform design throughout the process.

Our research design proceeded in two stages. First, we conducted pilot interviews (n=5) to refine the interview protocol and assess the sensitivity of the questions. We then carried out formal interviews (n=20), and once sufficient depth and variety of cases had been collected, we performed a theoretical saturation check.

\subsection{Participants}
Participants were primarily recruited from an online support community on the Chinese social media platform Xiaohongshu (“rednote”), where self-identified scam victims gather to share experiences and provide mutual aid. The research team posted recruitment messages in this community and also reached out via private messages to potential participants. After obtaining informed consent, we conducted semi-structured interviews through online video or audio conferencing tools. Each interview lasted between 60 and 90 minutes.

We finally recruited 25 participants, referred to as P1–P25 to preserve anonymity. The group included 11 males and 14 females, with ages ranging from 22 to 58 (median = 34). Their occupational backgrounds were diverse, including white-collar employees at internet companies (n = 7), freelancers (n = 5), small business owners (n = 6), retirees (n = 4), and graduate students (n = 3). The scams they encountered spanned multiple types, such as romance fraud (“pig butchering”), Pi cryptocurrency investment schemes, online gambling inducements, and debt-intermediary scams. About one-third of participants reported financial losses exceeding RMB 100,000, with several cases surpassing RMB 500,000, while others described scams involving smaller but more prolonged losses. Participants were located across major metropolitan areas in mainland China (e.g., Beijing, Shanghai, Guangzhou, Shenzhen), emerging “new first-tier” cities, as well as among overseas student communities. For ethical and privacy reasons, and in line with several participants’ explicit requests, we do not provide a detailed demographic table; background details are elaborated in the Findings section where relevant. All participants provided informed consent.

\subsection{Study Design}

Our study employed a semi-structured interview design to capture victims’ experiences with longer-running scams and the role of platform features in shaping these encounters. The design followed an iterative process in which we began with pilot interviews to refine the protocol and then conducted the main series of interviews using the revised guide. This iterative approach was particularly important given the sensitive nature of the topic: scam victimization often involves emotional distress and financial harm, and our study design needed to balance the pursuit of rich narratives with the ethical imperative to protect participants’ well-being.

\subsubsection{Pilot Interviews}

Before launching the main study, we conducted five 60-90 minutes pilot interviews to test the clarity, sensitivity, and sequencing of the interview protocol. These sessions revealed that some initial questions relied on academic terms such as “socio-technical affordances” or “platform affordances,” which participants found confusing or intimidating. Based on feedback, we rephrased these into more accessible, experience-oriented prompts, for example, asking\textit{ “What did the platform let you do that made you trust it more?”} or \textit{“How did the app encourage you to keep interacting with this person or group?” }The pilot interviews also helped us reconsider the order of topics. Instead of beginning directly with questions about financial losses—which risked triggering distress—we started with neutral questions about participants’ everyday social media use, then gradually transitioned into scam-related narratives. This sequencing was essential for reducing participant discomfort while also producing richer and more reflective accounts. In this way, the pilot stage was not simply a methodological exercise, but an ethical safeguard consistent with value-sensitive research practices: it enabled us to identify potentially harmful question framings, reduce emotional risk, and create a more supportive environment for participants.

\subsubsection{Formal Interviews}

Informed by the pilot stage, we then conducted twenty formal interviews using the revised protocol. Each interview lasted between 60 and 90 minutes and followed a structure organized into four thematic areas.

First, we explored participants’ initial exposure to scam-related content and the platform features through which this occurred. We asked participants to recount the first moment they encountered the scam, probing for the specific entry points such as recommendation feeds, search results, trending hashtags, or community groups. To understand perceptions of credibility, we asked what made the content or profile appear trustworthy—for example, whether it was a professional-looking avatar, the presence of a verification badge, a large number of comments or likes, or the apparent alignment with participants’ personal interests (such as finance, health, or lifestyle). Follow-up questions included: \textit{“What did you notice first about the post or profile?”} and \textit{“Did the way the platform presented this content affect how credible it looked?”}

Second, we examined the unfolding of the scam and key turning points in the interaction. Participants were encouraged to narrate how their relationship with the scammer developed over time, including the pacing and strategies of engagement. We asked them to describe major milestones such as when the scammer first suggested moving to a private chat, when they were invited into a group, when the first monetary transfer was requested, or when the scammer escalated to emotional appeals or promises of profit. We also probed for how trust was built, asking questions such as: \textit{“What did the scammer do or say that made you feel more comfortable?” }or \textit{“Were there moments when you had doubts, and how were those doubts addressed?”} This helped us map scams not as single interactions but as processes marked by critical decision points.

Third, we focused on the role of specific platform features and design elements in enabling trust and shaping engagement. Participants were asked to identify which features of the platform—such as algorithmic recommendations, payment systems, customer service chat, group chat, or comment sections—played a role in their experience. To probe deeper, we asked: \textit{“Did the design of the payment interface make you feel the transaction was safe?”} \textit{“How did verification badges, customer service responses, or system-generated records affect your perception of legitimacy?”} We also asked about exit difficulties: \textit{“Were there moments when you tried to withdraw or leave, and did the platform’s design make that harder?”} These reflections provided insight into how scams exploit the affordances and visual cues of mainstream platforms.

Finally, we asked participants for post-hoc reflections on their experiences and perceived causes of victimization. In this section, participants were encouraged to articulate their understanding of why they became involved in the scam. We asked: “Looking back, what do you think made you vulnerable in that situation?” and “Do you feel the platform or the scammer’s personal strategies were more influential in keeping you engaged?” To identify prevention opportunities, we also asked: \textit{“What might have helped you recognize or stop the scam earlier?”} and \textit{“If the platform had provided certain warnings, features, or support, what would have made the most difference for you?”} These reflective narratives not only highlighted individual sensemaking but also illuminated structural gaps in platform governance.

\subsection{Data Analysis}

All interviews were transcribed verbatim and, together with the supplementary artifacts (e.g., screenshots and transaction records), constituted the dataset for analysis. In total, we analyzed approximately 350,000 words of narrative material in Chinese and over 150 images of platform interactions. Our analytic approach was grounded in qualitative, interpretive methods, drawing particularly on reflexive thematic analysis \cite{braun2019reflecting}. 

The analysis proceeded in several stages. First, two members of the research team independently conducted open coding on a subset of five transcripts to identify recurrent patterns in participants’ accounts, paying special attention to how scams unfolded temporally and how platform features were implicated. This initial coding produced a shared codebook that combined inductive codes (e.g., “illusion of legitimacy,” “algorithmic nudges,” “group pressure”) with deductive sensitizing concepts from prior work on platform affordances and deceptive design.  

Second, using the agreed codebook, the full dataset was coded iteratively in NVivo. The coders met weekly to discuss discrepancies and refine the coding scheme. In these discussions, we emphasized interpretive depth over consensus coding: disagreements were treated as opportunities to explore alternative readings, which enriched rather than diluted the thematic development. Through this iterative process, we clustered codes into higher-level categories capturing the trajectory of scam experiences, including phases of exposure, trust building, coercion, and reflection.  

Third, we triangulated narrative accounts with the screenshots and transaction records that participants provided. These materials allowed us to verify specific details (e.g., the presence of verification badges, payment interface layouts, or the content of group chats) and to connect participants’ retrospective accounts with the concrete affordances of the platforms. This multimodal analysis helped us situate individual experiences within broader socio-technical contexts.  

Finally, we engaged in reflexive memoing throughout the analytic process to account for our own positionality and potential biases. Given the sensitivity of the topic, the team frequently revisited analytic decisions to ensure that themes were not only empirically grounded but also respectful of participants’ experiences. The resulting thematic structure captured both the micro-level tactics used by scammers and the macro-level platform features that sustained longer-running scams, forming the basis of the findings we report in the following section.

\section{Findings}
Our analysis of 25 victim narratives reveals how longer-running scams unfold not as isolated transactions but as socio-technical processes sustained by platform infrastructures. We identify three interrelated mechanisms that structure the scam lifecycle. First, scams gain their initial foothold through mainstream social media, where recommendation systems, content streams, and interactive spaces such as comment sections serve as entry points that lower suspicion and amplify exposure (\autoref{f1}). Second, once victims are drawn in, scammers strategically mobilize platform design features—including counterfeit trading interfaces, embedded payment systems, customer service chats, and group interactions—to fabricate a sense of legitimacy and transparency (\autoref{f2}). Third, scams are sustained over time through the relational and affective dimensions of platforms: group dynamics, emotional support, and disciplinary mechanisms bind victims into networks that are difficult to question and even harder to exit (\autoref{f3}). Together, these findings highlight how the very affordances that enable everyday sociality and participation on digital platforms can be repurposed to create durable infrastructures of fraud.

\subsection{Scam Entry Points: Tapping into Mainstream Social Media for Exposure and Amplification} \label{f1}

Our findings show that the first critical step in longer-running scams involves how potential victims are drawn into the fraudulent ecosystem. Unlike traditional scams that rely on cold calls or spam messages, these schemes often originate within mainstream social media platforms, gaining initial visibility through legitimate-seeming content feeds and interactive features. By framing fraudulent pitches as \textit{“experience sharing”} or \textit{“resource channels,”} scammers seamlessly embed their messages into the everyday rhythms of social platforms, allowing victims to enter scam trajectories with little initial suspicion.

In P2’s case, this process of initial credibility building was particularly salient. P2 first encountered scam-related content in the financial discussion section of Xiaohongshu. A self-styled \textit{“cryptocurrency investment expert” posted daily “live trading screenshots,”} showcasing profits in Bitcoin and Ethereum alongside professional-looking market commentary. Unlike unsolicited private messages, these posts appeared within the open, everyday environment of a popular social platform and were surfaced through algorithmic recommendations aligned with P2’s existing interest in finance. Within Xiaohongshu’s culture of \textit{“experience sharing,” }such posts blended seamlessly with other lifestyle or self-improvement content (e.g., fitness check-ins). As P2 recalled: \textit{“Everyone was sharing their daily experiences—he just posted his returns every day. It looked natural, so I thought it was trustworthy.”}

Over time, the recommendation algorithm amplified this exposure. Because P2 consistently engaged with financial content, the platform expanded his feed from general stock tips to riskier cryptocurrency investments and eventually leveraged derivatives. This \textit{“expansion effect”} not only ensured that scam content recurred in P2’s feed but also reshaped his sense of financial possibility, gradually pushing him toward more aggressive investments. As he explained: \textit{“At first, I only wanted to see stock advice. But after seeing so many posts about high returns, I started thinking I should try something more aggressive—maybe I could earn more too.”}

P7’s story further underscores how interactive features such as comment sections play a critical role in amplifying scam credibility. Searching Xiaohongshu for “personal credit” during a period of financial distress, P7 was directed to a post associated with the so-called \textit{“debt relief”} scam. The scheme lured struggling individuals with offers to forge documents, illegally obtain bank loans, and disappear with the funds. Initially, the fact that such content appeared on a legitimate platform reassured P7 that it was not suspicious. What ultimately solidified his perception of credibility, however, was the activity in the comment section.

There, purported \textit{“successful participants”} shared glowing testimonials, while numerous broker accounts openly solicited clients, leaving contact information and promising one-on-one guidance. This created an impression of an established \textit{“industry,”} normalizing the idea that debt scams were part of a legitimate financial ecosystem. As P7 recalled: \textit{“So many middlemen in the comments were competing for clients, saying they had resources and ways to help. I thought, if it were a scam, how could so many people openly advertise? It felt like a real business.”}

This dynamic had both social and technical consequences. Psychologically, the interactive space created a sense of communal endorsement and collective participation. Technically, the high volume of comments and engagements increased the post’s algorithmic weight, ensuring higher visibility for similar users. Thus, comment sections became a core part of the scammers’ strategy: by faking endorsements and mobilizing brokers, they simultaneously produced a \textit{“crowd-based legitimacy” }effect and exploited ranking algorithms to maximize exposure.

These cases highlight that scam entry points are not simply the result of direct persuasion by scammers but rather emerge through the intersection of scam tactics and platform architectures. The publicness and legitimacy of mainstream platforms lower initial suspicion, recommendation algorithms progressively steer users toward riskier topics, and interactive spaces like comment sections foster the illusion of community and industrialization. Importantly, the starting point of these scams is not hidden corners of the dark web but the highly visible, everyday infrastructures of mainstream social media. Under these layered mechanisms, victims’ initial trust is quietly but effectively constructed.

\subsection{Platform Design: The Fabrication of Legitimacy} \label{f2}
After victims are funneled into scam environments through mainstream platforms, scammers do not immediately ask them to transfer money or disclose sensitive information. Instead, they rely on a series of platformized design features to further construct credibility. These designs include both entirely fabricated trading interfaces and the strategic appropriation of legitimate features such as payment systems, customer service interactions, and group-based communication. Within this \textit{“authenticity manufactured by design,”} victims gradually lower their guard and become drawn into deeper relationships and financial transactions.

\subsubsection{Fake Exchange Interfaces}

Scammers first construct highly realistic virtual trading platforms to simulate the operating environment of mainstream financial applications. These interfaces typically include real-time market charts, order book depth tables, portfolio overviews, customer service pop-ups, and even notifications such as \textit{“new user registration successful.” }To enhance their credibility, many of these fake platforms introduce incentives like \textit{“zero commission for new users.”} For victims, such arrangements not only appear reasonable but also reinforce the impression that they are engaging with a genuinely operational platform.

P8’s experience illustrates the power of this tactic. After being funneled from Xiaohongshu, she clicked on a link to the scammer-recommended \textit{“exchange” }and found its interface even smoother than legitimate platforms she had used before. As she recalled:\textit{ “At that time, I truly believed it was a real trading platform—the page design was so convincing, even smoother than the official ones I had used before.” }She even saved screenshots of her “profit curves” and shared them with friends, further reinforcing her sense of authenticity. The deception worked not only by mimicking the visual aesthetics of financial trading but also by embedding operational details such as commission waivers and leaderboard updates, fabricating the illusion of a functioning market.

However, the interface alone is only the first step. To persuade victims to actually invest money, the design of the payment mechanism becomes the decisive factor.

\subsubsection{Payment Interfaces and Low-Threshold Transactions}

Many scam platforms lower victims’ psychological defenses by offering highly convenient payment interfaces. Unlike traditional bank transfers, these systems are often embedded directly into the platform and closely mirror familiar practices from mainstream applications such as Alipay or WeChat Pay, allowing for seamless \textit{“one-click transfers.”} This familiarity creates a sense of transparency and control, making victims more willing to experiment with small, trial investments.

Even more deceptively, platforms frequently encourage users to start with minimal amounts. Several victims described how their first deposits not only appeared to generate profits on the interface but also allowed them to successfully withdraw a small sum in the early stages. P3 recalled:\textit{ “I put in 100 yuan, and after two days it showed 120. I could actually withdraw the money. It felt so real—don’t regular platforms work the same way?”} This symbolic withdrawal instilled a powerful sense of trust, leading victims to believe the system operated according to genuine financial logic.

Once the initial step was taken, the platform guided users into gradually increasing their investments through tiered recharge mechanisms. Common tactics included unlocking \textit{“VIP strategies” o}nce deposits surpassed 10,000 RMB or restricting access to \textit{“special financial products” }to higher-balance accounts. This laddered design coerced users into repeatedly committing larger sums, caught between the fear of missing out on higher returns and the desire to protect existing investments. As the amounts escalated, victims increasingly found themselves trapped in a sunk cost dilemma, making it difficult to withdraw even when doubts emerged.

At the same time, the payment interface itself played a crucial role in justifying delayed withdrawals. When users attempted to withdraw larger sums, the system often displayed messages about \textit{“risk control reviews”} or \textit{“compliance verification,” }fabricating plausible reasons for delays. To reinforce this illusion, scammers sometimes deployed customer service representatives to intervene, offering professional-sounding explanations that reassured victims and dissuaded them from exiting. Several participants emphasized that because the interface maintained the appearance of compliance and professionalism, they were willing to wait rather than abandon their investments.

In short, payment interfaces functioned as more than transactional tools: they were carefully engineered mechanisms for manufacturing transparency and legitimacy. From the initial low-stakes trial and symbolic payout, to the laddered recharge system, to delayed withdrawals framed as regulatory checks, these design strategies collectively worked to reinforce victims’ trust and prolong their entanglement in the scam.

\subsubsection{Pseudo-Legitimate Customer Service Systems}
When victims began to feel uneasy due to delayed withdrawals or confusion about operations, the scam platform’s \textit{“customer service”} function typically intervened. These service windows were embedded directly into the trading interface in the form of a chat box, almost indistinguishable from those of mainstream e-commerce or financial applications. Victims could initiate conversations at any time and would receive prompt, polite replies. The customer service agents used professional and restrained language, frequently invoking terms such as \textit{“company policy,”} \textit{“risk notice,”} or \textit{“operational protocol,”} which reinforced the illusion of a regulated and institutionalized environment.

P12’s experience demonstrates the deceptive power of this mechanism. After depositing funds into an online gambling platform, he grew suspicious when his withdrawal was delayed and decided to contact the platform’s customer service. The representative responded immediately, offering a formal explanation: \textit{“Since the platform settlement cycle follows T+1, please wait patiently for the funds to arrive.” }The plausible technical language reassured him. As P12 recalled: \textit{“They spoke just like people at a bank—calm, professional, and with clear procedural explanations. I thought, if it were a scam, they wouldn’t explain things so carefully.”} In reality, these conversations were scripted in advance, designed precisely to rebuild trust at the critical moment when victims might otherwise disengage.

This shows how payment interfaces and customer service systems often worked in tandem: the former manufactured a sense of transparent process, while the latter stepped in to patch cracks whenever doubt surfaced. Together, they reinforced an infrastructure of false legitimacy that sustained victims’ continued involvement.

\subsubsection{Private Chats and Group Spaces}

Beyond interfaces and customer service interactions, scammers also leverage private messaging and group functions to construct sustained relational networks. Victims are often pulled into chat groups composed of \textit{“investment advisors,”} \textit{“successful students,”} and \textit{“ordinary investors,”} many of whom are in fact roles played by members of the scam team. These groups are highly active, with members posting daily \textit{“investment insights,” }profit screenshots, and market analyses, creating the impression of a thriving \textit{“investment community.”} Immersed in such interactions, victims gradually come to see themselves as part of a \textit{“genuine learning collective.” }As P21 recalled: \textit{“People in the group were discussing the market every day. It really looked like a real investment community—I even felt I had found a sense of belonging.”
}

This mechanism was particularly evident in P12’s case. He was initially added to a so-called \textit{“online gambling discussion group,”} where screenshots of recent winnings circulated constantly and administrators regularly distributed \textit{“newcomer benefits.” }More significantly, the gambling platform embedded a small commission mechanism into its rules: every bet or withdrawal carried a minimal transaction fee, often too small to arouse suspicion. For P12, this \textit{“reasonable commission” }reinforced the sense that the platform operated with institutional logic, resembling a regulated casino rather than a scam. As he explained: \textit{“At first I doubted it, but then I noticed every withdrawal had a tiny service fee. That made me believe it was real. After all, scammers just want to take all your money at once—they wouldn’t bother with small commissions.” }Such institutionalized details made the scam appear sustainable and credible in the eyes of participants.

Overall, scammers transform victims’ initial exposure on legitimate platforms into deeper trust through a combination of fabricated exchange interfaces, convenient payment channels, pseudo-legitimate customer service systems, and socialized group interactions reinforced by institutional details such as commissions. These mechanisms are not isolated: interface aesthetics establish the first impression, payment systems lower the entry threshold, customer service patches doubts at critical moments, and group spaces stabilize long-term relationships by cultivating community and legitimacy. Victims’ trust thus arises not purely from persuasive scammers but from the authenticity manufactured through platform design. In other words, the affordances of digital platforms are repurposed within scam contexts, becoming the very conditions that allow fraudulent schemes to sustain and evolve.

\subsection{Relationship Maintenance and Emotional Manipulation} \label{f3}

If funneling mechanisms and platform design provide the entry points and the appearance of legitimacy for scams, it is the interactive and social nature of platforms that supplies the soil for their long-term maintenance. Scams are not merely isolated financial operations; they are embedded within platform-based social functions, where interaction frequency, group belonging, and emotional investment gradually consolidate trust. In this environment, scammers often reap the benefits without exerting continuous pressure: platform logics of sociality and group atmosphere themselves cultivate trust and emotional attachment.

P4’s experience vividly illustrates the power of platform social functions in sustaining scams. A recently retired middle school teacher, P4 lived alone in an aging urban neighborhood. Following retirement, his daily social circle shrank—former colleagues were busy, neighborhood ties were distant, and his adult children worked far away. \textit{“Sometimes an entire day would pass without me talking to anyone. My phone rarely buzzed. By evening, it felt especially quiet,”} he recalled.

It was in this context of social isolation that P4 encountered a Pi Coin investment group. From his very first day in the chat group, he felt a long-missing sense of liveliness: members greeted each other with “good morning” messages, celebrated children’s college admissions, shared updates on their Pi Coin mining, and sent festive GIFs. Though the topics were mundane, this constant interaction made the group feel less like strangers and more like a dependable community.\textit{ “Gradually, I felt they weren’t just online acquaintances, but people I was living with. Even if Pi Coin was fake, I got used to chatting there,” }P4 said.

Over time, his investment became less about money and more about emotional dependency. Every morning upon waking, his first act was to check the hundreds of unread messages in the group. The lively discussions created a sense of presence and being needed. For P4, the continuous group activity compensated for loneliness stemming from absent children and cold neighborhood relations. The group chat became his primary site of social life.

The group atmosphere also created subtle disciplinary mechanisms. Certain members acted as \textit{“evangelists,”} consistently posting profit screenshots and defending the reputation of Pi Coin. In contrast, anyone expressing doubts was quickly mocked as \textit{“short-sighted”} or accused of lacking faith. P4 recalled witnessing such a scene: when someone questioned withdrawal delays, others retorted, \textit{“You only think about quick money, you don’t see the big picture.”} More intriguingly, whenever disputes escalated, scammers would intervene under the guise of \textit{“group leaders” }or \textit{“senior investors,” }urging members to \textit{“stay rational and united.”} This posture of mediation enhanced their image as responsible community managers rather than orchestrators of fraud.

Even after P4 began to suspect that Pi Coin was a scam, he hesitated to leave. \textit{“Leaving the group felt like cutting ties with friends. I rarely have people to talk to in daily life—if I closed this, I’d be completely alone,”} he explained. For him, the companionship and emotional support outweighed financial gains. Ultimately, his deepest entanglement was not the monetary loss but the exit barrier created by platform-based sociality: money could be forfeited, but relationships and belonging were far harder to sever.

Compared with P4’s deep emotional reliance, P11’s Pi Coin experience highlighted how platform-based communities also enforced discipline and exclusion. A self-employed shopkeeper, P11 first encountered Pi Coin through a friend’s forwarded message. He joined a chat group\textit{ “to learn more about an investment opportunity,” without fully believing at first. Months later, after experiencing repeated withdrawal delays, he voiced his doubts in the group: “Why do withdrawals always take so long? Is the platform short on funds?”}

This seemingly ordinary question immediately triggered a surge of hostile responses. Evangelists quickly jumped in to defend the scheme. Some accused him of being \textit{“short-sighted” }and caring only about quick money; others mocked him for having \textit{“weak faith”}; still others derided him as a “drag on the group.” Within minutes, the chat was flooded with attacks, and hardly anyone spoke in his defense.

What surprised him most was that the scammers themselves did not suppress him harshly but instead adopted the role of \textit{“neutral mediators.”} They stepped in to calm the group, urging members to \textit{“look at things rationally”} and to \textit{“not let outsiders sow division.”} For P11, this reinforced a deep sense of isolation: \textit{“All I wanted was clarity, but suddenly I was branded a traitor. Even the group owner was telling everyone to stay united. That’s when I knew I was alone.”}

Afterward, P11 stopped posting and remained a silent observer. He described himself as \textit{“lurking,”} reading the hundreds of messages that appeared each day without contributing. \textit{“Even though I knew something was wrong, I couldn’t stop checking. The group felt like an information window—if I left, I’d miss something important.” }This reveals the double logic of platform communities: on the one hand, continuous activity created a sense of presence and connection; on the other hand, collective discipline and peer pressure silenced dissent, yet kept skeptics tethered to the relational network.

Together, P4 and P11’s stories illustrate how the sociality of platforms sustains scams in two distinct ways. For older individuals lacking offline relationships, platform communities offer substitute emotional support, making them reluctant to leave even when suspicions arise. For questioning individuals, group dynamics amplify shame and exclusion, marginalizing dissenters while binding them through ongoing observation. Meanwhile, scammers position themselves as mediators rather than aggressors, further consolidating their authority in the group.

Overall, platforms were not simply channels through which scams spread but constituted the very social environments in which scams nested. Through visible group activity, amplified interactions, and continuous social hosting, platforms created networks that were difficult to exit and dangerous to question. It was within these environments that scams gained resilience beyond financial logic, becoming part of victims’ everyday routines and social lives.

\section{Discussion}

Our study shows that longer-running scams cannot be reduced to isolated acts of fraud or to individual lapses in judgment. Instead, they emerge through the interaction of malicious actors with the affordances, logics, and cultures of digital platforms. By examining scams as temporally extended and socio-technical projects, we move beyond explanations centered on user gullibility or after-the-fact detection. Our findings highlight three interlocking dynamics: (1) platform infrastructures that funnel users into risky content streams; (2) design features that fabricate legitimacy through interface conventions, payment flows, and scripted customer support; and (3) relational mechanisms that entangle victims in communities where exit is socially and emotionally costly. Together, these dynamics illustrate how scams exploit the very features that make platforms integral to everyday life, transforming tools of connection, personalization, and trust into infrastructures of deception. 

In what follows, we unpack these dynamics by situating our findings within broader HCI and platform studies scholarship, and by outlining how they challenge conventional approaches to online safety and platform governance. First, platforms function not as neutral backdrops but as socio-technical arenas where affordances such as recommendation systems, payment integrations, and private chats enable scammers to gradually draw users deeper into fraudulent networks (\autoref{dis1}). Second, scams unfold temporally and through layered mechanisms: exposure through algorithmic curation, legitimacy fabrication via interface and transaction design, and relational maintenance within community settings. These cumulative processes show why scams are resilient and why single-point interventions often fail (\autoref{dis2}). Third, these insights generate practical implications for design and governance. By embedding protective mechanisms into the architecture of exposure, transaction, and community features, platforms can shift from reactive enforcement toward proactive disruption of scam lifecycles (\autoref{dis3}). These three dimensions—platform arenas, layered temporality, and design implications—reframe scams not as isolated incidents of user vulnerability but as infrastructural challenges demanding systemic, design-centered responses.

\subsection{Platforms as Socio-Technical Arenas for Scamming} \label{dis1}
Our findings demonstrate that longer-running scams are not merely opportunistic acts by malicious actors but are deeply embedded within the sociotechnical infrastructure of platforms. Platforms should not be understood as neutral backdrops but as arenas that actively shape exposure, trust formation, and the sustainability of scams \cite{krebs2014spam,gillespie2010politics,helmond2015platformization}. Building on scholarship on platform affordances, we argue that features such as recommendation algorithms, payment systems, and private messaging afford scammers opportunities to gradually lead users from initial exposure to deep involvement. 

Recommendation and ranking systems determine what becomes visible to users, effectively creating a risk exposure architecture \cite{mcdonald2000expertise}. In our interviews, participants frequently attributed their first contact with scams to algorithmic recommendations that surfaced high-return investment posts, gradually leading them to riskier content such as cryptocurrency trading groups.
Moreover, our data illustrate how design choices can function as dark patterns that facilitate trust construction rather than manipulation for commercial gain. As HCI researchers note, dark patterns leverage psychological tendencies and interface expectations to push users toward actions they would not have otherwise taken \cite{monge2023defining,mathur2021makes,gray2018dark}. Scam interfaces exploited these mechanisms by mimicking legitimate financial apps, hiding critical information, and creating “roach motel” dynamics—easy entry but high exit barriers—mirroring manipulative retention tactics described in prior HCI research \cite{sheil2024staying}. These interface-level cues worked together with social proof signals (e.g., fake comments and testimonials) to create what our participants described as an “almost too-real” environment, which made disengagement emotionally and cognitively costly.

Viewing platforms as sociotechnical arenas thus reframes scams as co-constructed phenomena: the scammers’ strategies are amplified by algorithmic curation, legitimized by familiar UI conventions, and socially reinforced through community affordances. This perspective shifts attention away from blaming individual users’ gullibility and toward the structural conditions that make scams persistent and scalable. Future research and design interventions must account for this entanglement between platform infrastructure and malicious actors, considering not only what content is removed but also how platform affordances might be reconfigured to disrupt the stepwise accumulation of trust that sustains longer-running scams.

\subsection{Temporality and Layered Mechanisms in Longer-Running Scams} \label{dis2}

Our study reveals that longer-running scams are not single, instantaneous acts of deception but rather multi-stage, gradually unfolding processes. Victims’ trust is not established at a single moment but is progressively shaped through repeated interactions. We conceptualize this process as consisting of three interconnected layers—exposure, fabricated legitimacy, and relational maintenance—and argue that the cumulative effect of this chain is what sustains long-term scams.

The recommendation and ranking algorithms of mainstream social platforms determine the visibility of scam-related content at the outset. Unlike traditional scams that depend on cold contact, these schemes typically begin by aligning with users’ existing interests, gradually guiding them toward higher-risk information through algorithmic curation. For example, participant P2 initially browsed posts about stock trading experiences; as the platform learned his preferences, he was increasingly shown cryptocurrency trading updates and high-yield investment posts. This trajectory closely aligns with the “algorithmic rabbit hole effect \cite{kaiser2019implications},” in which recommendation systems progressively push users toward more extreme or attention-grabbing content, deepening immersion and engagement. In the scam context, this intensifying exposure is amplified by algorithmic reminders—similar to a “nagging pattern \cite{caragay2024beyond}”—that make it difficult for users to ignore such content, thereby directing them toward the scam’s entry point.

Once victims enter the scam environment, they encounter highly realistic trading interfaces, seamless payment integrations, and responsive customer service—design elements that together create a powerful sense of operational realism. Previous HCI studies show dark patterns frequently exploit users’ familiarity with interfaces and their psychological expectations to guide behavior \cite{gray2021end}. Scam platforms mimic the norms of legitimate financial applications, obscure risk warnings, and streamline deposit processes, reducing victims’ suspicion. Moreover, many participants reported being able to withdraw small amounts after their initial deposits, generating positive reinforcement. This, in turn, activated a sunk cost effect, encouraging progressively larger investments. We conceptualize this layer as a fusion of the “obstruction pattern” and “roach motel” design: entry is easy but exit is difficult, producing a one-way accumulation of trust and financial commitment.

As investments increased, victims were drawn into group chats, private messaging channels, or so-called “learner communities,” where they received both emotional support and investment advice. The key mechanisms here were social proof and community norm enforcement. Groups were filled with screenshots of profits, success stories, and enthusiastic endorsements, reinforcing the collective belief that “everyone is making money,” and thereby lowering individual thresholds for doubt. At the same time, skeptics were often ridiculed or ostracized, creating powerful normative pressure that forced them either into silence or out of the group altogether. Scammers often reinforced their authority by adopting the stance of neutral mediators, calming disputes and encouraging unity. As participant P4 put it, \textit{“Leaving the group felt like cutting ties with friends,”} underscoring how relational maintenance functioned not only as emotional support but also as social locking, where exit incurred both identity and emotional costs.

This three-layer model demonstrates the cumulative nature of scams: algorithmic recommendations produce persistent exposure, interface design and early positive feedback establish fabricated legitimacy, and social interactions embed scams into everyday relationships, ultimately making it difficult for victims to disengage. The cumulative effect also suggests that one-off interventions (e.g., a single pop-up warning) are unlikely to disrupt the closed loop of the scam, since risk perception is repeatedly recalibrated across interactions. Accordingly, protective measures must adopt a staged approach: visibility warnings at the exposure stage, cooling-off periods and multi-factor verification at the interface layer, and low-cost exit mechanisms and external support channels at the community level, in order to interrupt the progression from initial contact to deep entrapment.

\section{Future Implications on Platform Design} \label{dis3}

Our study not only reveals how scams strategically exploit platform features but also offers new directions for platform governance and interaction design. Existing protective measures often emphasize account suspension or warning pop-ups after scams are detected, but they overlook the temporality and multi-layered mechanisms of scam progression. Building on our three-layer model, we suggest the following design and governance strategies.

First, at the stage of exposure, platforms should enhance risk visibility and algorithmic transparency. Recommendation and ranking systems could provide explicit risk signals and credibility markers for high-risk financial content. For example, when users repeatedly encounter posts about high-return investments or cryptocurrency trading within a short period, the recommendation system could trigger explanatory prompts clarifying that such content may be unverified or risky. This design aligns with the principle of algorithmic interpretability \cite{eslami2018communicating,eslami2016first}, helping users recognize when they are being guided toward a high-risk exposure pathway. Platforms could also disclose portions of recommendation logic or risk assessment indicators to improve user understanding of algorithmic choices and reduce the blind trust created by “black box” opacity.

Second, during the stage of legitimacy fabrication, platforms should embed trust friction and introduce tiered verification. Within trading interfaces and payment flows, this could include mechanisms such as secondary confirmations, cooling-off periods, or multi-factor authentication when users attempt large transactions or multiple rapid transfers. Such interventions create “windows for reflection” and disrupt impulsive decisions made under heightened arousal. Platforms could also impose stricter identity verification on in-house exchanges or investment tools and mandate public disclosure of operator information, reducing the likelihood that fraudulent interfaces masquerade as legitimate financial applications.

Third, at the stage of relational maintenance, platforms should lower exit costs and strengthen support channels. For group chats and community spaces, this could mean designing one-click exit options that simultaneously block related recommendations, allowing users to sever ties with scam-related groups more easily. Embedded reporting and support channels should enable users to directly flag suspicious communities to platforms or third-party agencies. To prevent victims from feeling socially isolated, platforms could also implement conversational interventions—delivering educational resources or psychological support when users attempt to leave scam groups or halt payments—thereby mitigating the emotional pressure associated with exit.

\section{Limitations and Future Work}

While our study provides new insights into how longer-running scams exploit platform infrastructures, several limitations warrant attention. First, our empirical material is drawn from 25 victims in China. Although this context offers valuable perspectives given the integration of financial and social functions within Chinese platforms, the findings may not generalize to other regions or platform ecosystems. Comparative studies across different cultural and regulatory contexts, especially within the Global South, could reveal how variations in platform design and governance shape scam practices and user vulnerabilities.

Second, our study relies primarily on retrospective self-reports from victims. Although we supplemented interviews with screenshots and transaction records to enhance validity, participants’ narratives may still be influenced by memory biases or post-hoc rationalizations. Future work could triangulate these accounts with data from platform audits, law enforcement cases, or computational analyses of scam-related content circulation.

Third, our research focuses on the victim side of the scam lifecycle. We did not have access to scammers themselves, nor to internal platform data about detection and moderation. Gaining visibility into scammer strategies and platform governance practices would provide a more holistic understanding of how scams evolve and persist. Partnerships with platforms, NGOs, or regulators may be necessary to access such perspectives.

Finally, our findings suggest that scams unfold through temporally layered mechanisms, but our qualitative approach limits our ability to quantify their prevalence or to test the effectiveness of specific interventions. Future work could employ longitudinal experiments or mixed-methods approaches to evaluate how staged interventions, such as algorithmic risk prompts, trust-friction designs, or community exit supports, can disrupt the cumulative entrapment of victims.  

\section{Conclusion}

This paper has examined how longer-running scams exploit the infrastructures and affordances of digital platforms to draw victims in, fabricate legitimacy, and sustain relationships over time. We showed that victims are rarely reached through random encounters; rather, they are funneled through mainstream platforms, where recommendation systems and interaction features amplify exposure and normalize fraudulent content. We demonstrated how scammers strategically mimic and repurpose the design logics of legitimate platforms, from polished trading interfaces and seamless payment flows to community spaces that provide emotional support, to cultivate trust and entrap victims.  

By conceptualizing scams as socio-technical projects, our study shifts attention away from explanations that emphasize individual gullibility or purely technical detection. Instead, it foregrounds the layered temporality of scams and the ways in which platform affordances themselves can be reconfigured into infrastructures of fraud. This perspective carries implications not only for HCI and online safety scholarship but also for platform governance and policy, highlighting the need to embed protective mechanisms into the very architectures of recommendation, payment, and community interaction.  

Ultimately, our work underscores that combating scams requires systemic approaches that move beyond reactive account suspension or user education. To effectively disrupt longer-running scams, interventions must address the socio-technical environments in which scams thrive, introducing transparency, friction, and support at multiple stages of the scam lifecycle. Thus, we argue for a reframing of scams as infrastructural challenges, entangled with the very features that make platforms integral to everyday life, and for future efforts that treat safety not as an afterthought, but as a core design principle of digital platforms.

\bibliographystyle{ACM-Reference-Format}
\bibliography{references}

\end{document}